\documentclass[12pt]{article}
\usepackage{times}
\usepackage{geometry}
\usepackage{epsfig, epstopdf}
\usepackage[utf8]{inputenc}
\usepackage{enumitem,amssymb}
\usepackage{ragged2e}
\newlist{thematic}{itemize}{8}
\setlist[thematic]{label=$\square$}
\usepackage{pifont}
\usepackage{amsmath}
\usepackage{amssymb}

\newcommand{\simgt}{\lower 2pt \hbox{$\, \buildrel {\scriptstyle >}\over {\scriptstyle\sim}\,$}}
\newcommand{\simlt}{\lower 2pt \hbox{$\, \buildrel {\scriptstyle <}\over {\scriptstyle\sim}\,$}}

\begin{document}

\def\sarc{$^{\prime\prime}\!\!.$}
\def\arcsec{$^{\prime\prime}$}
\def\beginrefer{\section*{References}%
\begin{quotation}\mbox{}\par}
\def\refer#1\par{{\setlength{\parindent}{-\leftmargin}\indent#1\par}}
\def\endrefer{\end{quotation}}

{\raggedright
\huge
Astro2020 Science White Paper \linebreak
\\
\noindent
A New Era for X-ray Lensing Studies of Quasars and Galaxies  \linebreak
\normalsize

\noindent \textbf{Thematic Areas:} \hspace*{60pt}  $\square$ Planetary Systems \hspace*{10pt} $\square$ Star and Planet Formation \hspace*{20pt}\linebreak
$\text{\rlap{$\checkmark$}}\square$ Formation and Evolution of Compact Objects \hspace*{31pt} $\text{\rlap{$\checkmark$}}\square$ Cosmology and Fundamental Physics \linebreak
  $\square$  Stars and Stellar Evolution \hspace*{1pt} $\square$ Resolved Stellar Populations and their Environments \hspace*{40pt} \linebreak
  $\text{\rlap{$\checkmark$}}\square$    Galaxy Evolution   \hspace*{45pt}  $\text{\rlap{$\checkmark$}}\square$     Multi-Messenger Astronomy and Astrophysics \hspace*{65pt} \linebreak

\textbf{Principal Author:}

Name: George Chartas	
 \linebreak						
Institution:  Department of Physics and Astronomy, College of Charleston
 \linebreak
Email:  chartasg@cofc.edu
 \linebreak
Phone:  814-441-4127
 \linebreak
 
\textbf{Co-authors:} Henric Krawczynski (Washington University in St. Louis), David Pooley (Trinity University), Richard F. Mushotzky (University of Maryland, College Park), Andrew J. Ptak (NASA, GSFC)
  \linebreak

\textbf{Abstract:}}
Current X-ray observations and simulations show that gravitational lensing can be used to infer the structure near the event horizons of black holes, constrain the dynamics and evolution of black-hole accretion and outflows, test general relativity in the strong-gravity regime and place constraints on the evolution of dark matter in the lensing galaxies. These science goals currently cannot be achieved in a statistically large sample of $z$ = 0.5 $-$ 5 lensed quasars due to the limited capabilities of current X-ray telescopes and the relatively low number ($\sim$ 200) of known lensed quasars. The latter limitation will be resolved with the multi-band and wide-field photometric optical survey of {\sl LSST} that is expected to lead to the discovery of $>$ 4,000 additional gravitationally lensed systems. As we show in this white paper, these science goals can be reached with an X-ray telescope having a spatial resolution of  $\simlt$~0\sarc5 to resolve the lensed images and a collecting area of $\simgt$ 0.5 m$^{2}$ at 1~keV.

\pagebreak
\vspace{-3mm}
\section{Introduction}
\vspace{-3mm}
The spectra and images of gravitationally lensed quasars contain a wealth of information on both the quasar and the foreground lens galaxy. Macrolensing is the gravitational bending of light produced by the global mass distribution of the lensing galaxy. This lensing is responsible for the production of multiple images (Figure 1), image flux ratios, lensing magnification and time delays. Microlensing is the gravitational bending of light produced by stars in the lensing galaxy (e.g., Paczynski 1986; Wambsganss 1990). Microlensing will lead to additional magnification of the affected images.

We identify the following scientific opportunities and compelling scientific themes for the coming decade, that have arisen 
from recent advances in X-ray optics and detector  technology and the expected synergy between a high-spatial resolution X-ray mission with the next generation of astronomical observatories (such as {\sl JWST, WFIRST, LSST, SKA, eROSITA, ATHENA, TMT, E-ELT, GMT, or CTA}):

\noindent
\begin{itemize}
\item Structure and evolution of AGN environments.

\item Measuring the dark matter fraction of the lensing galaxies as a function of redshift that will provide insight to the evolution of galaxies.


\item Measuring  SMBH spin and its evolution over cosmic time that will provide insight to the accretion and merger history of AGN.

\item Cosmic feedback from quasar winds at the peak of AGN activity.
\end{itemize}

\vspace{-7mm}
\section{Structure and evolution of AGN environments}
\vspace{-3mm}
Kochanek (2004) described a microlensing light-curve method that compares the variability amplitudes in the X-ray and optical of individual lensed images to place tight constraints on the structure of the accretion disk and the X-ray emitting regions. 
An application of this microlensing light-curve method to {\it Chandra} observations of several bright lensed quasars reveal that the X-ray emitting corona is very compact, $\sim 6-50 r_{g}$ ($r_{g} = GM_{BH}/c^{2}$; Pooley 2009; Chartas et al. 2016).  The compactness of the hot corona has been independently confirmed with X-ray reverberation mapping studies of nearby Seyfert galaxies (Cackett et al. 2014, Kara et al. 2016).

The microlensing light-curve method is currently limited to the relatively small number of known gravitationally lensed systems and the imaging and sensitivity limitations of {\sl Chandra}.  One of these limitations will soon be lifted with the aid of multi-band and wide-field photometric optical surveys of the Large Synoptic Survey Telescope ({\it LSST}),  
{\sl GAIA}, and {\sl EUCLID}. The {\it LSST} survey ($\sim$ 2023$-$2033) alone is expected to lead to the discovery of $>$4,000 additional gravitationally lensed quasars (e.g., {\it LSST} Science Book; Oguri and Marshall 2010). In Figure 1 we show the angular separation of lensed quasars expected to be discovered by {\it LSST}. A high-resolution and high-throughput X-ray mission (FWHM $\simlt$ 0\sarc5), however, will be required to resolve all of the newly discovered {\it LSST} (FWHM$\sim$0\sarc7) lensed quasars. The application of the microlensing light-curve method to the {\it LSST} lensed quasars over a range of quasar redshifts, black hole masses, radio loudness values, and Eddington ratios will constrain the dependence of the sizes of X-ray emitting regions over this parameter space.

The application of the microlensing light-curve method to a high signal-to-noise (S/N $\simgt$ 200) X-ray spectrum of a lensed quasar will also constrain the sizes of X-ray emitting regions of quasars ranging from the hot corona and inner accretion flow to the molecular and dusty torus. Isolating these emission regions is possible in a high S/N  X-ray spectrum because the microlensing light-curve method can be applied to select energy bands over which these emission components dominate.  
{\it LSST} will be monitoring the fluxes of the images of lensed quasars over its ten-year survey period and can provide robust triggers for high-resolution and high-throughput X-ray missions to cover single caustic crossing events.
The sample of newly discovered {\it LSST} lensed quasars available for simultaneous monitoring in optical and X-ray bands will be two orders of magnitude larger than the lensed systems currently known. We estimate, however, that an effective area of an order of magnitude larger than 
{\sl Chandra} would be required to detect $\sim$300 quadruply lensed and $\sim$1000 double quasars X-ray bright enough to employ the microlensing techniques discussed. The X-ray fluxes of the newly discovered {\sl LSST} lensed quasars will be determined from {\sl eROSITA} measurements. {\sl eROSITA} is the primary instrument onboard the Russian/German ``Spectrum-Roentgen-Gamma'' satellite scheduled to be launched in $\sim$2019 $-$2020. During its mission it will perform an all sky survey and will provide the X-ray fluxes of the newly discovered {\sl LSST} lensed quasars.

\begin{figure}
    \centering
    \vspace*{-10mm}
  \includegraphics[width=.8\textwidth]{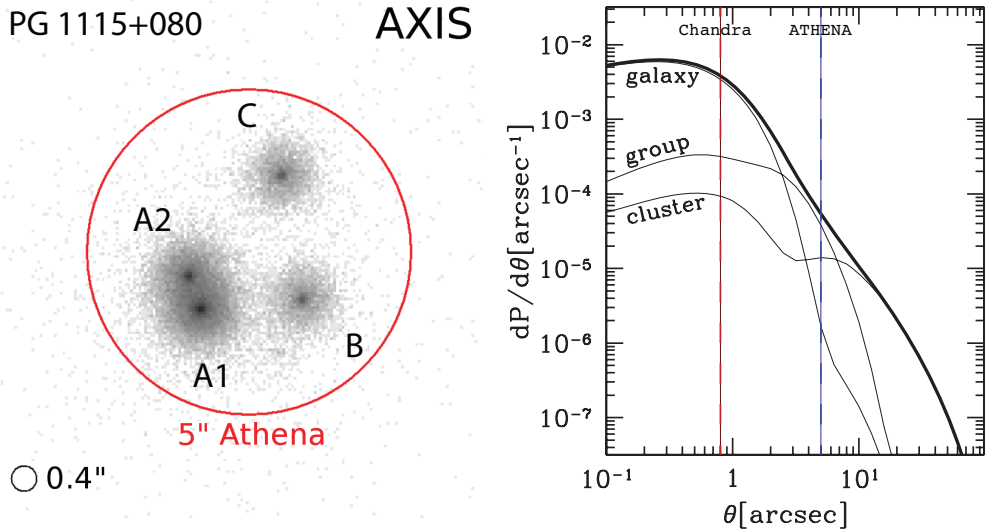}
     \caption{ (Left)  A sub-arcsec mission like {\sl AXIS} or {\sl Lynx} will be able to measure the spin of the black hole in the lensed quasar PG1115+080  through monitoring variability of the quasar's multiple lensed images.  A 30 ks AXIS exposure (0\sarc4 HPD) will yield high-quality spectra from each quasar image and the variability on timescales much shorter
than those accessible for {\sl Chandra}.  (Right) The distribution of lens image separations for three different scales: galaxy, group, and cluster-scales, predicted by a halo model (Oguri 2006). The total distribution is shown by the thick line. The vertical lines represent the half power diameters of {\sl Chandra} and {\sl Athena}.}
  \vspace*{-4.mm}
    \label{fig:lens_image_separations}
\end{figure}

\vspace{-3mm}
\section{Measuring the evolution of dark matter fraction of galaxies}
\vspace{-3mm}
The main parameters of a microlensing model include the sizes of the emission regions, the average mass of the stars doing the microlensing, the mean surface density in the stars of the lensing galaxy, the fraction of normal matter to dark matter in the lensing galaxy and the velocity describing the motion of the AGN regions across the microlensing caustics.  Simulations of light-curves of images of lensed quasars from caustic crossings are fit to observed light-curves to constrain these parameters. 
The success of the microlensing light-curve method has been demonstrated with optical ground based, {\it HST}, and {\it Chandra} observations of several bright lensed quasars (e.g., Morgan et al. 2008; Chartas et al. 2009; Dai et al. 2010; Morgan et al. 2010; Mosquera et al. 2013; Blackburne et al. 2014,2015; MacLeod et al. 2015). 

An independent method for estimating the different fractional contributions of stars and dark matter to the total surface density was developed by Schechter \& Wambsganss (2002,2004).
They found that the probability of a strong demagnification of a saddle point image, was relatively low for stellar fractions of 2\% and 100\% but became appreciable for stellar fractions of 5\%$-$25\%. The application of this method to X-ray observations of lensed quasars resulted in tight constraints on the dark matter fraction of the lensing galaxies (Pooley et al. 2012).

The large sample of lenses that can be simultaneously monitored with a next-generation, sub-arcsecond  X-ray mission and  {\sl LSST} will allow us to determine the evolution of the properties of galaxies at $0.1 < z < 2$.

\begin{figure}[t]
    \centering
       \vspace*{-2mm}
  \includegraphics[width=0.9\textwidth]{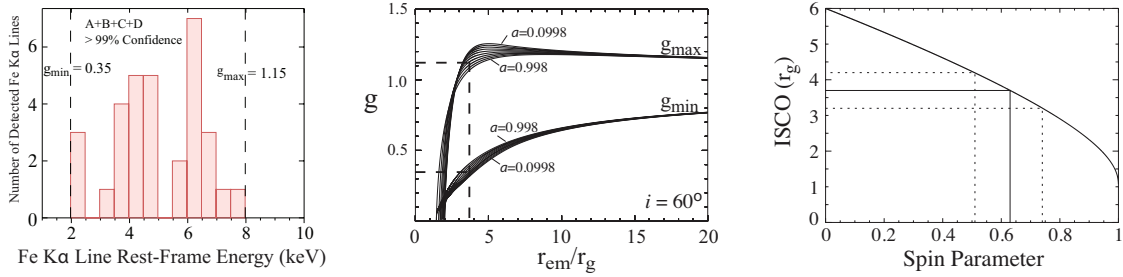}
     \caption{(Left) Distribution of the Fe K$\alpha$ energy-shifts for all {\it Chandra} images for RXJ1131. Only cases where the iron line is detected at $>$  99\% are shown. (Middle) Simulated extremal shifts of the Fe K$\alpha$ line energy for spin values ranging between 0.1 and 0.998 in increments of 0.1 for an inclination angle of $i$  = 60~degrees . The horizontal lines show the extreme $g = E_{\rm obs}/E_{\rm rest}$ values for Fe K$\alpha$ lines detected at $>$ 99\%. (Right) The $g$-distribution constrains the ISCO of RXJ1131 to be 3.7 $\pm$ 0.5 $r_g$ that corresponds to a spin parameter of $a$ = 0.63 $\pm$ 0.10}
        \vspace*{-3mm}
    \label{fig:lens_g_distribution.}
\end{figure}

\vspace{-3mm}
\section{Measuring SMBH spin and its evolution}
\vspace{-3mm}
The mechanisms leading to the growth of supermassive black holes can be studied by constraining the evolution of the SMBH spins of lensed quasars with redshifts in the range of  0.5 $-$ 5.  
Numerical simulations indicate that the evolution of the spin of supermassive black holes will depend on the accretion rate, the mode of accretion and mergers. Specifically, in the high redshift universe, numerical simulations predict that accretion rates onto the central SMBHs of galaxies are high and so the spin evolution is dominated by the high accretion. At low redshifts, galaxies are in general more massive and gas poor and SMBH$-$SMBH mergers are predicted to lead to the decrease of SMBH spins (e.g., Dubois et al. 2014; Volonteri et al. 2013).  
There are two microlensing methods that can be applied to X-ray observations of lensed quasars to measure their spin and ISCO sizes over the $z=0.5-5$ redshift range and thus strongly constrain numerical models of black hole growth  (Volonteri et al., 2013). \\

\noindent
{\it Method 1: $g$-distribution method}\\
Quasar microlensing produces energy shifts of the Fe~K line emitted from the accretion disk (Chartas el al. 2017; Krawczynski \& Chartas 2017; Krawczynski et al. 2019).These shifts originate from general and special relativistic effects when the magnification caustic selectively magnifies emission very close to the event horizon of the black hole. The distribution of Fe K$\alpha$ energy-shifts ($g$-distribution) for $z$=0.658 quasar RX J1131$-$1231 is shown in Figure 2, where $g = E_{obs}/E_{rest}$ is the fractional energy shift of the iron line. By comparing the distribution of energy shifts (g-distribution method) to simulated ones we constrain the ISCO, spin, and inclination angle of distant quasars (see Figure 2).  To study the evolution of ISCO and spin of quasars over the $z$ = 0.5 $-$ 5 redshift range one would monitor a sample of $\sim$100 X-ray bright quadruply lensed quasars over this redshift range. For each quasar we would build a $g$-distribution to infer the ISCO and spin. \\

\begin{figure}
    \centering
  \includegraphics[width=0.9\textwidth]{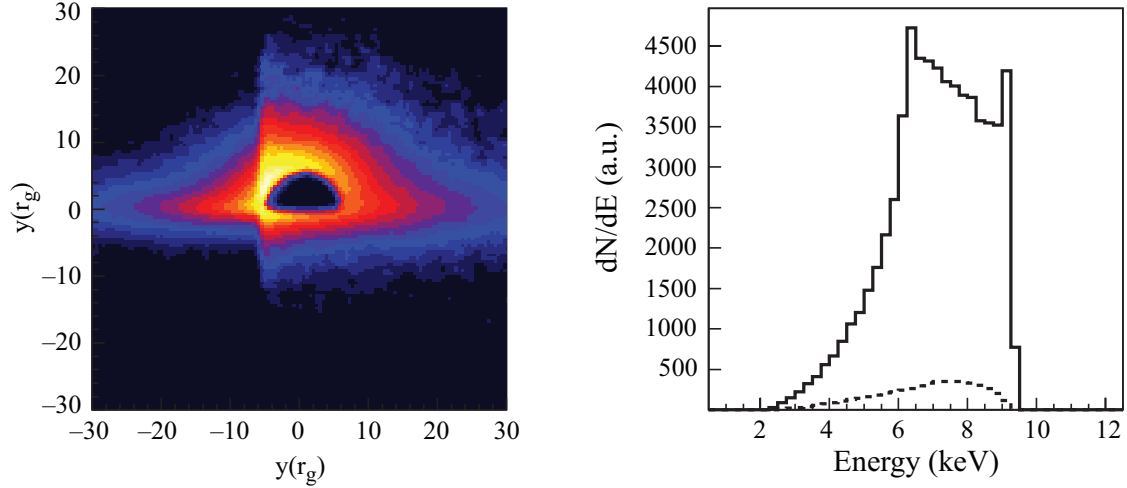}
     \caption{The left panel shows an image of the surface brightness of the Fe-K$\alpha$ line emission as seen by a distant observer and an inclination of $i$  = 82.5 degrees from a black hole of spin $a$ = 0.3. The right panel shows the resulting energy spectrum of the Fe-K$\alpha$ emission in the rest frame of the source. The dashed line shows the Fe K line with no microlensing present.}
    \label{fig:lens_simulated_crossing}
\end{figure}

\noindent
{\it Method 2: Caustic Crossing Method}\\
The detection of individual caustic crossing events would be spectacular revealing the gradual change in the profile and energy of the Fe line as the caustic sweeps over the accretion disk and corona (see simulation in Figure 3).  Monitoring caustic crossings events in individual images with a high-spatial resolution X-ray mission can provide tomographic scans of the accretion disks of supermassive black holes.
Having an X-ray mission that combines high spatial resolution and effective area that overlaps with {\it LSST} is important: Since {\it LSST} will be continuously monitoring the lensed quasars it can provide reliable triggers for caustic crossing events.
Neronov \& Vovk (2016) show that the observed time-dependent shifts produced in a single caustic crossing event can also be used to test the Kerr metric of a supermassive black hole in the strong-gravity regime (see their Figure 5 and their Equation 16).

The microlensing method is independent of the fitting procedure for the standard relativistic iron line method (e.g., Fabian et al. 1989; Laor 1991; Reynolds \& Nowak 2003) of determining spin in unlensed quasars.  Both methods make different assumptions for estimating spin and it is important to measure the spin of black holes using independent tests to confirm the reliability of spin measurements. 

\vspace{-3mm}
\section{Feedback from quasar winds at the peak of AGN activity}
\vspace{-3mm}
Relativistic wide-angle outflows of AGN are now considered one of the main mechanisms regulating the
evolution of galaxies through a feedback process (e.g., see review by King \& Pounds (2015), and references within). These wide-angle winds are thought to transfer a substantial amount of
their kinetic energy to the surrounding gas resulting in quenching of star formation in the host galaxy by
heating the interstellar medium or by ejecting the gas from the galaxy 
(e.g., Faucher-Gigu{\`e}re, C.-A., \& Quataert, E., 2012; Zubovas, K., \& King, A. 2012).

The detections of ultrafast outflows in distant quasars are rare due to their X-ray weakness. 
The first evidence of relativistic winds in quasars came from observations of APM~08279$+$5255, PG~1211+143, and PDS~456 
(Chartas et al.\ 2002; Pounds et al.\ 2003; Reeves et al. 2003).
Follow-up studies of a larger sample of nearby Seyfert galaxies showed that about 40\% of these AGN have highly-ionized ultrafast outflows (UFOs)
with velocities exceeding 10,000 km~s$^{-1}$ and with average velocities ranging between 0.1$c$ and 0.3$c$ (Tombesi et al.\ 2010; Gofford et al.\ 2013).
There have been attempts to compare the energetics of small scale ultrafast outflows with larger scale molecular outflows in galaxies to test feedback models. The presence of both small and large scale energy-conserving outflows were recently discovered by Tombesi et al. (2015), 
in the $z$~=~0.189 ULIRG IRAS F11119+3257, 
and by Feruglio et al.\ (2015), in the $z = 0.04217$ ULIRG Mrk~231. 
Feruglio et al.\  (2017), have also reported the detection of molecular gas outflowing with maximum velocity of $v$~=~1,340~km~s$^{-1}$ in the $z$ = 3.912 BAL quasar APM~08279+5255.

Studying ultrafast outflows in gravitationally lensed quasars provides a significant advantage due to the lensing magnification that typically can range between 5$-$100.  As a result, most of the known quasars with relativistic outflows are gravitationally lensed.
With more than  4,000 new lensed quasars to be discovered with {\sl LSST}, {\sl GAIA} and {\sl EUCLID} and with {\sl eROSITA} to provide the X-ray fluxes of these quasars there is an important opportunity in the next decade to study the feedback process in a large sample of AGN spanning a range of quasar redshifts, black hole masses, radio loudness values, and Eddington ratios.

A high spatial resolution and high throughput X-ray mission in the next decade would be required to resolve the spectra of the individual lensed images.
Such a mission would provide the spatially resolved and time-resolved spectra of lensed images, thus constraining the properties of the outflow in individual images. Lensed images provide spectra of the quasar at different epochs separated by the image time delays. Detecting the acceleration phase of the outflowing absorber and constraining the short timescale variability of the properties of the absorber cannot be accomplished with longer exposure times using current X-ray missions but requires a mission with a significantly larger collecting area

The spectral-timing analysis of a sample of high magnification lensed high-$z$ quasars will constrain the energetics of ultrafast outflows near the peak of AGN activity. A comparison between the energetics of these small-scale ultrafast outflows and the large scale molecular outflows in the host galaxies would be used to infer their contribution to regulating the evolution of their host galaxies (e.g., Tombesi et al. 2015; Feruglio et al. 2015, 2017).

\pagebreak
\noindent
{\bf \Large References}\\
\\
\noindent
Blackburne, J.~A., Kochanek, C.~S., Chen, B., Dai, X., \& Chartas, G.\ 2014, ApJ, 789, 125 \\
Blackburne, J.~A., Kochanek, C.~S., Chen, B., Dai, X., \& Chartas, G.\ 2015, ApJ, 798, 95 \\ 
Cackett, E.~M., Zoghbi, A., Reynolds, C., et al.\ 2014, MNRAS, 438, 2980 \\
Chartas, G., Brandt, W.~N., Gallagher, S.~C., \& Garmire, G.~P.\ 2002, ApJ, 579, 169\\
Chartas, G., Kochanek, C.~S., Dai, X., Poindexter, S., \& Garmire, G.\ 2009, ApJ, 693, 174 \\
Chartas, G., Rhea, C., Kochanek, C., et al.\ 2016, Astronomische Nachrichten, 337, 356  \\
Chartas, G., Krawczynski, H., Zalesky, L., et al.\ 2017, ApJ, 837, 26 \\
Dai, X., Kochanek, C.~S., Chartas, G., et al.\ 2010, ApJ, 709, 278 \\
Dubois, Y., Volonteri, M., \& Silk, J.\ 2014, MNRAS, 440, 1590 \\
Fabian, A.~C., Rees, M.~J., Stella, L., \& White, N.~E.\ 1989, MNRAS, 238, 729 \\
Faucher-Gigu{\`e}re, C.-A., \& Quataert, E.\ 2012, MNRAS, 425, 605  \\
Feruglio, C., Fiore, F., Carniani, S., et al.\ 2015, A\&A, 583, A99 \\
Feruglio, C., Ferrara, A., Bischetti, M., et al.\ 2017, A\&A, 608, A30 \\
Gofford, J., Reeves, J.~N., Tombesi, F., et al.\ 2013, MNRAS, 430, 60 \\
Kara, E., Alston, W.~N., Fabian, A.~C., et al.\ 2016, MNRAS, 462, 511 \\
King, A., \& Pounds, K.\ 2015, ARA\&A, 53, 115 \\
Kochanek, C.~S.\ 2004, ApJ, 605, 58 \\
Krawczynski, H., \& Chartas, G.\ 2017, ApJ, 843, 118 \\
Krawczynski, H., Chartas, G., \& Kislat, F.\ 2019, ApJ, 870, 125 \\
Laor, A.\ 1991, ApJ, 376, 90 \\
LSST Science Collaboration, Abell, P.~A., Allison, J., et al.\ 2009, arXiv:0912.0201 \\
MacLeod, C.~L., Morgan, C.~W., Mosquera, A., et al.\ 2015, ApJ, 806, 258 \\
Morgan, C.~W.; Kochanek, C.~S.; Dai, X.; Morgan, N.~D.; Falco, E.~E. 2008, ApJ, 689, 755.\\
Morgan, C.~W., Kochanek, C.~S., Morgan, N.~D., \& Falco, E.~E.\ 2010, ApJ, 712, 1129 \\
Mosquera, A.~M., \& Kochanek, C.~S.\ 2011, ApJ, 738, 96 \\ 
Mosquera, A.~M., Kochanek, C.~S., Chen, B., et al.\ 2013, ApJ, 769, 53 \\
Neronov, A., \& Vovk, I.\ 2016, Physical Review D, 93, 023006 \\
Oguri, M.\ 2006, MNRAS, 367, 1241
Oguri, M., \& Marshall, P.~J.\ 2010, MNRAS, 405, 2579 \\
Paczynski, B.\ 1986, ApJ, 301, 503 \\
Pooley, D., Rappaport, S., Blackburne, J.~A., et al.\ 2009, ApJ, 697, 1892 \\
Pooley, D., Rappaport, S., Blackburne, J.~A., et al.\ 2012, ApJ, 744, 111 \\
Pounds, K.~A., Reeves, J.~N., King, A.~R., et al.\ 2003, MNRAS, 345, 705 \\
Reeves, J.~N., O'Brien, P.~T., \& Ward, M.~J.\ 2003, ApJL, 593, L65 \\
Reynolds, C.~S., \& Nowak, M.~A.\ 2003, Physics Report, 377, 389 \\
Schechter, P.~L., \& Wambsganss, J.\ 2002, ApJ, 580, 685 \\
Schechter, P.~L., \& Wambsganss, J.\ 2004, Dark Matter in Galaxies, 220, 103 \\
Tombesi, F., Cappi, M., Reeves, J.~N., et al.\ 2010, A\&A, 521, A57 \\
Tombesi, F., Mel{\'e}ndez, M., Veilleux, S., et al.\ 2015, Nature, 519, 436 \\ 
Volonteri, M., Sikora, M., Lasota, J.-P., \& Merloni, A.\ 2013, ApJ, 775, 94 \\
Wambsganss, J.\ 1990, Ph.D.~Thesis \\
Zubovas, K., \& King, A.\ 2012, ApJL, 745, L34 \\

\end{document}